\documentclass[onecolumn,11pt]{article}

\usepackage{graphics}
\usepackage[pdftex]{graphicx}
\usepackage{amssymb}
\usepackage{amsmath}
\usepackage{amsthm}
\usepackage{hyperref}
\begin{document}

\title{Approximate Bayesian Solution for Estimating Population Size from Dual-record System}

 \author{Kiranmoy Chatterjee\thanks{Department of Statistics, Bidhannagar College, Kolkata-700064, India; E-mail: \emph{kiranmoy07@gmail.com}} \\
Diganta Mukherjee
\thanks{Indian Statistical Institute, Kolkata-700108, India.}}

\date{}
 \maketitle

\begin{abstract}
 For Dual-record system, in the context of human population, the popular Chandrasekar-Deming model incorporates only the time variation effect on capture probabilities. However, in practice population may undergo behavioral change after being captured first time. In this paper we focus on the Dual-record system model (equivalent to capture-recapture model with two sampling occasions) with both the time as well as behavioral response variation. The relevant model suffers from identifiability problem. Two approaches are proposed from which approximate Bayes estimates can be obtained using very simple Gibbs sampling strategies. We explore the features of our two proposed methods and their usages depending on the availability (or non-availability) of the information on the nature of behavioral response effect. Extensive simulation studies are carried out to evaluate their performances and compare with few available approaches. Finally, a real data application is provided to the model and the methods.

\paragraph{}\emph{Key words:} Approximate Bayesian Computing, Behavioral response effect; Homogeneous Human population; Identifiability; Model $M_{tb}$.
\end{abstract}

\section{Introduction}
\label{sec:intro}

Estimation of coverage error in census operation is an important statistical and demographical issue. Federal agencies are generally interested to know the actual size of a specified population or any vital event occurring in a specified area within a given time span. Census or any vital events registration system often fails to extract the actual size. As a remedy, one or more information is collected on that population independently, near to the census operation, and the population size (say, \emph{N}) is estimated by matching the available lists of information. In the context of human population, this kind of data structure by matching information is known as Multiple-record system, which is equivalent to the capture-recapture system in biological or epidemiological studies (\textit{see} Otis et al., 1978\cite{Otis78} and Seber, 1986\cite{Seber86}). Now a days this technique is increasingly employed to correct for under ascertainment in traditional epidemiological surveillance (Chao et al., 1996\cite{Chao96}, 2001\cite{Chao01}). In dual-record system (or shortly, DRS), two samples ($T=2$) and in triple-record system (TRS), three samples ($T=3$) of information are used to estimate the \emph{N}. Undercount estimation in census (Wolter, 1986\cite{Wolter86}; Cressie, 1989\cite{Cressie89}; Wang et al., 2006\cite{Wang06}), extent of registration for vital events (Chandrasekar et al., 1949\cite{Chadra49}; Nour, 1982\cite{Nour82}), estimation of the size of drug abused population (Xu et al., 2014\cite{Xu14}), etc., are the primal potential application of DRS model. In connection with undercount rate estimation in US Census, different capture-recapture models based on DRS have been reviewed by Wolter (1986\cite{Wolter86}). The most simple and widely practiced model is $M_t$, which accounts for only time (\textit{t}) variation effect, was first invoked in modern times by Chandrasekar et al. (1949\cite{Chadra49}) on human population. There are various frequentist and likelihood approaches for this basic model in the literature of animal population size estimation (\textit{see}, Huggins (1989\cite{Hugg89})) and epidemiology from capture-recapture experiment (\textit{see}, Hook (1995\cite{Hook95}) and Chao et al. (2001\cite{Chao01})). Bayesian approach was pioneered by Robert (1967\cite{Robert67}), Castledine (1981\cite{Castle81}) and Smith (1988\cite{Smith88}), primarily for $M_t$. George and Robert (1992\cite{George92}) first gave an extensive account on the population size estimation through hierarchical Bayesian analysis on model $M_t$.

But this common DSE model doesn$'$t work well because of the violation of the underlying assumption of independence between the capture probabilities in different samples. Brittain and B$\ddot{o}$hning (2009\cite{Brittain09}) efficiently reviewed various methods available by relaxing the independence assumption and associated comparative study undertaken in the DRS context. For a homogeneous population, the capture probabilities might become correlated due to list dependence and it occurs for the time ordered samples when capture probability at the time of second survey depends on whether he/she is captured in first time. This behavioral dependency is driven by a parameter $\phi$ ($\in R^{+}$), called \emph{Behavioral Response Effect}. When behavior response (\textit{b}) effect acts together with the time variation (\textit{t}) effect on the capture probability in second list, then we would have a more complicated model $M_{tb}$. This model has a strong relevance in practice for a group of homogeneous individuals when the sample lists are not thought to be independent.
Otis et al. (1978\cite{Otis78}) addressed the identifiability problem related to the model $M_{tb}$. In recent past, Link (2003), Holzmann et al. (2006) and Farcomeni and Tardella (2012) worked on identifiability issues for heterogeneous population. However, in this article, we confine ourselves to a homogeneous population. Lloyd (1994\cite{Lloyd94}) used a martingle approach to solve the problem using an assumption in case the number of capturing ($T$) is strictly more than two. Quasi-likelihood method by Chao et al. (2000\cite{Chao00}) and univariate Markovian approach proposed by Yang and Chao (2005\cite{Yang05}) also successfully solve the nonidentifiability for $T\geq3$ and provide significant improvement over the classical solutions - unconditional and conditional MLE obtained from the popular assumption of Lloyd (1994\cite{Lloyd94}). In Bayesian paradigm, Lee and Chen (1998\cite{Lee98}) applied the Gibbs sampling idea to the model $M_{tb}$ but they did not use recapture data and estimates were unstable and prior sensitive. Lee et al. (2003\cite{Lee03}) considered the model with an informative uniform prior on $\phi$ and came up with a fully Bayesian solution using adaptive acceptance-rejection sampling. To discover a reasonable range for $\phi$, they require large number of samples (i.e. $T\geq4$) likely for animal population size estimation but unlikely for human. In a DRS framework, Nour (1982\cite{Nour82}) proposed an estimator assuming the directional knowledge of recapture proneness (\textit{i.e.} $\phi>1$) in human demographic studies. Another possibility is that recapture aversion (\textit{i.e.} $\phi<1$) might take place in some situations, e.g. drug abused population size estimation. The goal of this article is to investigate the potential of an approximate Bayesian methodology for homogeneous human population size ($N$) estimation as an alternative or supplement to the very few existing approaches when behavioral effect plays a key role along with time variation.

In this article we confine ourselves only to DRS, as three or more sources of data is seldom used for human population. In the next section, first we present the data structure in DRS and discuss the model $M_{tb}$.
It is clear from the nature of the present model that without the help of informative prior, it is not possible to have a reasonably good Bayesian solution. Hence, we design approximate Bayesian strategy using vague prior based on available directional knowledge on $\phi$. By understanding a need to develop an approach when no such directional knowledge on $\phi$ is available, another Bayesian strategy is proposed using a subjective choice of conjugate prior.
In both situations, priors specifications using the available domain knowledge, preserve some characteristics of the system and use minimal subjectivity so that efficient solutions could be obtained simply. Both the strategies are presented along with extensive numerical illustration over several simulated populations in section \ref{sub:first methodology_combo} and \ref{sub:second methodology_combo}. The full Bayes strategy is also presented without consideration of the directional knowledge as developed in Lee et al. (2003\cite{Lee03}). Section \ref{sec:real data} examines a real DRS data on death records as an illustrative example. Finally in the last section, we summarize our findings and provide the best possible estimation rule depending upon the availability (or non-availability) of the directional knowledge on $\phi$.

\section{Dual-record System: Preliminaries}
\label{sec:dse}

The idea of dual collection is equivalent to the very popular capture-recapture sampling in wildlife management to estimate the population size, $N$. Laplace (1786) pioneered this kind of sampling to estimate the population size of France from vital events like births, marriages and deaths. Let us consider a human population \emph{U} of size \emph{N}. Census usually fails to count this $N$ exactly and therefore, it can be considered as a sample (though a very large one for human population). To know the true \emph{N}, another independent attempt is organized. Then combining these two sources of information, estimate of \emph{N} could be obtained assuming different conditions on the capture probabilities of both the lists leading to different models. In this article we will concentrate on the situation which has two basic assumptions - (\textit{i}) population is closed upto the time of second sample taken, (\textit{ii}) individuals are homogeneous with respect to their capture probabilities. Let, $p_{j,1\cdot}$ and $p_{j,\cdot1}$ denote the capture probabilities of \emph{j}th individual in first sample (List 1) and second sample (List 2) respectively. Hence, homogeneity assumption in (\textit{ii}) ensures that $p_{j,1\cdot}=p_{1\cdot}$ in List 1 and $p_{j,\cdot1}=p_{\cdot1}$ in List 2 for \emph{j}=1, 2, $\cdots$, \emph{N}. Some countries use their regular survey as the second source, e.g. US Census Bureau uses the Current Population Survey(CPS) to estimate the true population. The individuals captured in first list (i.e. census) are matched one-by-one with the list of individuals made from second survey (often called Post Enumeration Survey or PES). Hence the data structure classified according to a multinomial fashion as in Table $\ref{tab:DSE}$ is popularly known as Dual-record system or shortly, DRS. The number of missed individual by both systems (denoted as $x_{00}$) is unknown which makes the total population size \emph{N} unknown. Expected Proportions or probabilities for each multinomial cell are also given in Table $\ref{tab:DSE}$ and these notations will be followed throughout this paper.
\begin{table}[h]
\footnotesize
\centering
\caption{$2\times2$ data structure for Dual-record-System (DRS) with probability of each cell mentioned in [ ] and $p_{\cdot\cdot}$=1}
\begin{tabular}{lccc}
&\multicolumn{3}{c}{List 2} \\
\cline{2-4}
List 1 & In & out & Total\\
\hline \hline
&\multicolumn{3}{c}{I. Observed sample numbers} \\
In & $x_{11}[p_{11}]$ & $x_{10}[p_{10}]$ & $x_{1\cdot}[p_{1\cdot}]$\\
Out& $x_{01}[p_{01}]$ & $x_{00}[p_{00}]$ & $x_{0\cdot}[p_{0\cdot}]$\\ 
\hline
Total& $x_{\cdot1}[p_{\cdot1}]$ & $x_{\cdot0}[p_{\cdot0}]$ & $x_{\cdot\cdot}=N[p_{\cdot\cdot}]$\\
\hline
\end{tabular}
\label{tab:DSE}
\end{table}
Usually, independence between the two lists is assumed to estimate the $N$ and that gives $\hat{N}_{ind}=(x_{\cdot1}x_{1\cdot}/x_{11})$. This is a popular estimate used in different contexts for homogeneous human population.
\subsection{Model $M_{tb}$}
\label{sub:complex model}

Consider a situation for a homogeneous population with the assumptions (\textit{i}) $p_{1\cdot}\neq p_{\cdot1}$ and (\textit{ii}) $p_{\cdot1}\neq p$, where $p=$ Prob(An individual is captured in List 2 $|$ he/she is not captured in List 1). Assumption (\textit{ii}) refers a violation of causal list-independence. Hence, the model becomes complex as it suffers from identifiability problem for drawing inference on $N$, as addressed by Otis et al. (1978\cite{Otis78}). We re-parameterize the recapture probability, $c=$ Prob(An individual is captured in List 2 $|$ he/she is not captured in List 1), as equals to a constant multiple of $p$, say, $c=\phi p$. Here $\phi$ represents the \textit{behavioral response effect} and likelihood becomes
\begin{equation}\label{likeli:Mtb}
L_{tb}(N,p_{1\cdot},p,\phi|D) \propto \frac{N!}{(N-x_0)!}\phi^{x_{11}}p_{1\cdot}^{x_{1\cdot}}p^{x_{\cdot1}}(1-p_{1\cdot})^{N-x_{1\cdot}}(1-p)^{N-x_{0}}(1-\phi p)^{x_{10}}.
\end{equation}
Thus, in a population, individuals are said to be \emph{recapture prone} if $\phi>1$ and they are called \emph{recapture averse} if underlying $\phi<1$.\\
\\
\textit{Remark 1:} In particular, when $\phi=1\Leftrightarrow c=p\Leftrightarrow$ causal independence holds, then (\ref{likeli:Mtb}) reduces to
\[L_{t}(N,p_{1\cdot},p_{\cdot1}) \propto \frac{N!}{x_{11}!x_{01}!x_{10}!(N-x_0)!}p_{1\cdot}^{x_{1\cdot}}p_{\cdot1}^{x_{\cdot1}}(1-p_{1\cdot})^{N-x_{1\cdot}}(1-p_{\cdot1})^{N-x_{\cdot1}}\]
as conditional $p$ will be equal to the unconditional $p_{\cdot1}$. Hence, this likelihood reduces to the model $M_t$ (Wolter, 1986\cite{Wolter86}) which is very simple and widely used for human populations under the capture-recapture framework. Associated MLE of $N$ exists and it is $\hat{N}_{M_t}=(x_{\cdot1}x_{1\cdot}/x_{11})=\hat{N}_{ind}$. But often in practice, this estimate is seriously criticized for the underlying causal independence assumption.\\
\\
\textit{Remark 2:} When $p_{1\cdot}=p_{\cdot1}=p^*$ (say) in $M_{tb}$, that is when there is no \textit{time variation effect}, another model $M_b$ is to be found (Wolter, 1986\cite{Wolter86}). The associated likelihood will be
\[L_{b}(N,p^*,c) \propto \frac{N!}{(N-x_0)!}p^{*x_{0}}c^{x_{11}}(1-p^*)^{2N-x_0-x_{1\cdot}}(1-c)^{x_{10}}.\]
Then, MLE $\hat{N}_{M_b}=x_0/\left\{1-\left(\frac{x_{01}}{x_{1\cdot}}\right)^2\right\}$.\\
\\
\textit{Remark 3:} Nour(1982\cite{Nour82}) considers situation of model $M_{tb}$ with $\phi>1$, i.e., \textit{recapture prone} population, as it is often likely in demographic study of human population. However, theirs approach is not model based and estimate is $\hat{N}_{Nour}=x_0+\frac{2x_{11}x_{10}x_{01}}{(x_{11}^2+x_{10}x_{01})}$.

\section{Solution with Available Directional Knowledge on $\phi$}\label{sub:first methodology_combo}
\subsection{Proposed Methodology (AB-Flat)}
\label{sub:first methodology}

In this section, we suggest an approximate Bayesian strategy and proposed associated computation technique so that the unidentifiability burden can be successfully overcome for estimating $N$ with noninformative flat prior depending on the directional knowledge on $\phi$. In (\ref{likeli:Mtb}), $\phi$ and $p$ both are not identifiable but their product $c$ ($=\phi p$) is identifiable and well estimated by $\hat{c}_{mle}=x_{11}/x_{1\cdot}$. Hence, for given $c$, we consider $p$ as a function of $\phi$ only as $p=p(\phi|c)=c/\phi$. Now we assign independent priors on $\Theta=(N, \phi, p_{1\cdot})$ as $\pi(\Theta)=\pi(p_{1\cdot})\pi(\phi)\pi(N)$ except $p$. Hence, separate conditional posterior distributions for $\Theta$ are obtained as follows
\begin{eqnarray}
\pi(p_{1\cdot}|N) & \propto & p_{1\cdot}^{x_{1\cdot}}(1-p_{1\cdot})^{N-x_{1\cdot}}\pi(p_{1\cdot}),\label{eqn:cond_post_p_1.}\\
\pi(\phi|N,p) & \propto & \phi^{x_{11}}(1-\phi p)^{x_{10}}\pi(\phi),\label{eqn:cond_post_phi}\\
\pi(N-x_0|p_{1\cdot},p) & \propto & \frac{N!}{(N-x_0)!}((1-p_{1\cdot})(1-p))^{N}\pi(N),\label{eqn:cond_post_N}
\end{eqnarray}
In addition to this, Lee et al. (2003\cite{Lee03}) also takes into account a prior on $p$; so they have identified known conditional posterior densities for $\textit{N}$, $p_{1\cdot}$ and employed adaptive rejection sampling to generate $p$ and $\phi$ as explicit conditional posteriors for $p$ and $\phi$ are not available.

We consider the noninformative prior for $p_{1\cdot}$ as $\pi(p_{1\cdot})=\emph{Unif}(0, 1)$. A flat prior density $\pi(\phi)=\emph{Unif}(\alpha, \beta)$ is chosen for $\phi$. It follows that (\ref{eqn:cond_post_p_1.}) and (\ref{eqn:cond_post_phi}) reduce to
\begin{eqnarray}
\pi(p_{1\cdot}|N) &\propto & Beta(x_{1\cdot}+1,N-x_{1\cdot}+1),\label{eqn:cond_post_p_1._2}\\
\pi(\phi|p) &\propto & \mbox{GB-I}(x_{11}+1,x_{10}+1,1,rate=p)\times \mathcal{I}_{[\alpha,\beta]}(\phi),\label{eqn:cond_post_phi_2}
\end{eqnarray}
where $\mathcal{I}_{[\alpha,\beta]}(\phi)$ is an indicator function for $\phi \in [\alpha,\beta]$ and GB-I refers \textit{Generalized Beta Type-I} density. Now, the hyperparameters $\alpha$ and $\beta$ are to be chosen. If no other information on $\phi$ is available then choosing prior distribution is not at all easy. Lee et al. (2003\cite{Lee03}) proposed a trial-and-error procedure for this. They opt for such $\alpha$ and $\beta$ for which the range of the posterior credible interval for $\phi$ is not too close to either side of the prior limits. They mentioned that this procedure seems to work well only when there is a large amount of recapture information. They also agreed that such kind of trial-and-error method has no theoretical justification and this judgement is highly subjective. For human population, high capture probabilities may be attained but number of samples is too small, usually, \textit{T}=2. Now, we propose a model-driven prior limits for $\pi(\phi)$, or can say it is rather an automatic choice. Since $c=\phi p<\phi$, \emph{c} is a good choice for the lower limit $\alpha$. Hence, our selected objective prior is $\pi(\phi)=Unif(c,1)$ when we know $\phi<1$. If it is known that the population is recapture prone (i.e. $\phi>1$), we recommend to set $\alpha$ to 1 and upper limit $\beta=2$ or $3$ is suitable for the analysis of human population.
When directional knowledge is not at all available, $\alpha=c$ and $\beta=2$ or $3$ is a safe choice. Thus, $\pi(\phi)$ would be non-informative irrespective of the availability of directional knowledge on $\phi$.
Now, two different priors on $N$ are considered as follows:

\emph{I.} Poisson prior: $\pi(N)=Poi(\lambda)$, then conditional posterior (\ref{eqn:cond_post_N}) becomes
\[ \pi(N-x_0|p_{1\cdot},p)\propto Pois(\lambda(1-p_{1\cdot})(1-p)) \mbox{\hspace{0.1in} and}\]

\emph{II.} Jeffrey's prior: $\pi(N)\propto 1/N $, then conditional posterior (\ref{eqn:cond_post_N}) becomes
\[ \pi(N-x_0|p_{1\cdot},p)\propto NegBinom(x_0,\mu),\]
where, $\mu$=$1-(1-p_{1\cdot})(1-p)$ and $p=c/\phi$, from definition of $\phi$. For poisson prior, we can use empirical estimate of $\lambda$, as stated in George and Robert (1992\cite{George92}). Here, we replace $\lambda$ by $\hat{N}_{M_b}$ (\textit{see} Remark 2, section \textbf{2}). Jeffrey's prior on \emph{N} is also equivalent to the prior $\pi(N)=Poi(\lambda)$ with $\pi(\lambda)\propto 1/\lambda$. For the case $\phi>1$, we also judge the performance of $\lambda=\hat{N}_{Nour}$.

At first we fix initial values $N^{(0)}$ and $\phi^{(0)}$. Then the initial value $p^{(0)}$ is obtained easily from the parametric relation $p=\hat{c}/\phi$ as $\hat{c}$ is consistent estimate for $c$. One can generate $p_{1\cdot}^{(0)}$ from the conditional posterior (\ref{eqn:cond_post_p_1._2}) replacing $N$ by $N^{(0)}$. Thereafter, Gibbs sampler, a Markov Chain Monte Carlo (MCMC) method proceeds as follows to obtain approximate posterior, especially for $N$.
\\
\\
\texttt{Step 1: Simulate $\phi^{(1)}$ from $\pi(\phi|p^{(0)})$, in (\ref{eqn:cond_post_phi_2}).\\
Step 2: Simulate $N^{(1)}$ from $\pi(N-x_0|p_{1\cdot}^{(0)},\phi^{(1)})$ and obtain $p^{(1)}=\hat{c}/\phi^{(1)}$.\\
Step 3: Generate $p_{1\cdot}^{(1)}$ from $\pi(p_{1\cdot}|N^{(1)})$, in (\ref{eqn:cond_post_p_1._2}).\\
Step 4: Repeat the above three steps until the convergence.} \\
\\
Thus, the above approach produces a Gibbs sequence \{$N^{(h)}$, $\phi^{(h)}$, $p^{(h)}$, $p_{1\cdot}^{(h)}$; \emph{h} = 1, 2, 3,...\} by repeating this process 2k ($k>0$, is to be specified) times. The initial value for $\Theta$ comes from a wide range of choices, so it is generally unstable at beginning of the process. To avoid the influence of the starting value, we discard the first \emph{k} iterative values as under the burn-in period and consider the remaining consecutive values to construct approximate posterior distributions for the model parameters. Here $p$ is simulated from degenerate density (or equivalently, from conditional density $\pi(p|\phi)$ with point-mass prior) at $p=\hat{c}/\phi$, conditional on $\phi$ and flat prior choice on $\phi$ is made based on available domain knowledge. Thus, we call this Bayesian strategy as approximate Bayes with flat prior on $\phi$ and denote it as \textit{AB-Flat}. We think such kind of degenerate prior satisfying a structural relation, will help to get rid of from the model complexity, especially when, one try to avoid subjective prior. Hence, The above approach is presented as a potential alternative to $N$ estimation problem under the model $M_{tb}$ when only two samples are available. Advantage of AB-Flat over Lee et al. (2003\cite{Lee03}) is that the computational burden can be successfully overcome in order to generate Gibbs samples for $p$. Another advantage is that we don't need to setup prior limits for $\phi$ by trial-and-error method. Another relation, $p=x_{01}/(N-x_{1\cdot})$, suggested by Llyod (1994\cite{Lloyd94}), can also be used in Step 2 in lieu of $p=\hat{c}/\phi^{(1)}$. There are many methods available for diagnostic checking of posterior convergence in literature. In this study, we use the iterative simulation technique using multiple sequence method (\textit{see} Gelman, 1996\cite{Gelman96}) and compute $\hat{R}^{1/2}$ exactly following Lee et al. (2003\cite{Lee03}, p.p. 483).
\subsection{Numerical Illustrations}
\label{sub:first illustration}

In this section we evaluate the performance of our proposed approach and understand its efficiency in order to apply the method when directional knowledge on $\phi$ is available. Let us simulate eight hypothetical populations corresponding to four pairs of capture probabilities ($p_{1\cdot}, p_{\cdot1}$)=\{(0.50, 0.65), (0.60, 0.70), (0.80, 0.70), (0.70, 0.55)\} in each case of \emph{recapture prone} (represented by $\phi=1.25$) and \emph{recapture averse} (represented by $\phi=0.80$) situations. All the populations are truly of size $N=500$. We denote four populations corresponding to four pairs of ($p_{1\cdot}, p_{\cdot1}$) with $\phi=1.25$ as P1, P2, P3 and P4 respectively and associated results are presented in Table $\ref{tab:recapture_prone}$. Results of another four populations, correspond to $\phi=0.80$ namely as P5, P6, P7, P8, are shown in Table $\ref{tab:recapture_aversion}$. The expected number of distinct captured individuals ($E(x_0)$) in each population is also cited in the respective Tables $\ref{tab:recapture_prone}$ and $\ref{tab:recapture_aversion}$. First two cases with $p_{1\cdot}<p_{\cdot1}$, represent the usual situation in DRS from Post Enumeration Survey (PES) conducted for estimating census undercount estimation. The hyper-parameters $\alpha$ and $\beta$ for $\pi(\phi) \propto U(\alpha,\beta)$ are taken as
\[ \mbox{($\alpha,\beta$)=($1,2$) for prior domain knowledge $\phi>1$ and}\]
\[ \mbox{($\alpha,\beta$)=($\hat{c},1$) for prior domain knowledge $\phi<1$.}\]
The above priors are chosen depending upon the available true directional knowledge on $\phi$. In addition to that, if no directional information on $\phi$ is available, another prior $U(\hat{c},2)$ is chosen assuming $\phi<2$. 200 data sets on ($x_{1\cdot},x_{\cdot1},x_{11}$) are generated from each of the eight populations. Our AB-Flat estimates have been obtained through simple Gibbs sampling from five independent parallel chains. Burn-in period is fixed at \emph{k} = 2000 in general, after observing the performance of $\hat{R}^{1/2}$. Finally, estimate of $N$ is obtained by averaging over 200 posterior means. Based on those 200 estimates, the bootstrap sample s.e. and sample RMSE (Root Mean Square Error) are calculated. We also compute the $95\%$ credible interval (C.I.) based on sample quantile of the approximate posterior distribution of \emph{N}. In addition to that, to compare the performance of our proposed AB-Flat method with a full Bayesian strategy developed by Lee et al. (2003\cite{Lee03}), we compute the similar statistics. However, Lee et al. (2003\cite{Lee03}) illustrated their approach in the context of animal capture-recapture experiment with a large number of sampling occasions. Theirs detail computation strategy, particularly for DRS, can be found in Chatterjee and Mukherjee (2014\cite{Chatterjee14}). Nour's (1982\cite{Nour82}) estimates are also calculated only for $\phi=1.25$ cases as, Nour (1982\cite{Nour82}) deduced his approach for \textit{recapture prone} situation. Theirs estimates as well as its S.E., RMSE, $95\%$ confidence interval are computed over 200 generated datasets and present them as average estimate, sample SE, sample RMSE, $95\%$ CI respectively in Table $\ref{tab:recapture_prone}$. When it is known that underlying $\phi>1$, we also evaluate the performance of our AB-Flat approach with $\pi(N)\propto Poi(\lambda=\hat{N}_{Nour})$ (\textit{see} Remark 3, in section \textbf{2}).

Population P1 and P2 in Table $\ref{tab:recapture_prone}$ demonstrate populations with $p_{1\cdot}<p_{\cdot1}$. Here we evaluate the performance of our proposed approach with Jeffrey's prior (in first row), poisson prior with $\lambda=\hat{N}_{M_b}$ (in second row) on $N$ corresponding to each of the four priors on $\phi$ previously mentioned. Estimates from priors $U(1, 2)$ or $U(1, 3)$ on $\phi$ are compared with the available Nour's (1982) estimator. Results suggest that when one has the information that $\phi>1$, the uniform priors $U(1, 2)$ or $U(1, 3)$ for $\phi$ is recommended for use. Another estimator is produced with prior Pois($\lambda=\hat{N}_{Nour}$) (in third row) for \textit{N}, where $\hat{N}_{Nour}$ is the estimate of $N$ due to Nour (1982 \cite{Nour82}). Tables show that the proposed approach from these two recommended priors are significantly better than the Nour's estimate based on RMSE and tighter confidence intervals around the true \emph{N}. Results from Jeffrey's or Pois($\lambda=\hat{N}_{Nour}$) prior on \emph{N} is better than that from Pois($\lambda=\hat{N}_{M_b}$). Overall, our proposed approximate Bayes solution performs very well and shows improvement over both the Lee's and Nour's approach (available only for $\phi>1$). On the contrary, when this knowledge is not available, the prior $U(\hat{c},2)$ performs not well unless the chance to be captured in List 1 is very high. This happens as $U(\hat{c},2)$ extends the constant prior below 1 up to $\hat{c}$, so that estimates become negatively biased.
\begin{table}[ht]
\footnotesize
\begin{center}
\caption{Summary results reflecting the performances of the proposed approach and other relevant methods applied to the populations P1-P4. Upper panel refers the situation when the knowledge, $\phi>1$, is available. Lower panel refers when nothing is available on $\phi$.}
\begin{minipage}{13cm}
\begin{tabular}{|lllcccc|}
\hline
Population& & Prior & Average & Sample & Sample &  \\
(E($x_0$)) & Method: $\pi(\phi)$ & $\pi(N)$ & Estimate & SE & RMSE & $95\% CI$ \\
\hline
\hline
\multicolumn{7}{|c|}{when it is known that $\phi>1$}\\
P1(394)&Nour& - & 479 & 14.84 & 25.49 & (452, 509)\\
&Lee: U(1,2)&Jeffrey & 472 & 18.94 & 34.11 & (438, 518) \\
&AB-Flat: U(1,2)&Jeffrey & 497 & 20.51 & 20.89 & (459, 539) \\
& &Poi($\hat{N}_{M_b}$)& 520 & 29.73 & 35.92 & (469, 582) \\
&  &Poi($\hat{N}_{Nour}$)& 489 & 17.99 & 21.04 & (456, 526) \\
P2(422)&Nour& - & 487 & 13.18 & 18.47 & (461, 512)\\
&Lee: U(1,2)&Jeffrey & 478 & 11.04 & 24.33 & (451, 519) \\
&AB-Flat: U(1,2)&Jeffrey & 491 & 15.36 & 17.97 & (460, 521) \\
& &Poi($\hat{N}_{M_b}$)& 492 & 15.84 & 17.77 & (460, 524) \\
&  &Poi($\hat{N}_{Nour}$)& 488 & 14.57 & 18.78 & (459, 517) \\
P3(458)&Nour& - & 499 & 8.74 & 8.76 & (481, 516)\\
&Lee: U(1,2)&Jeffrey & 490 & 7.55 & 12.02 & (473, 515) \\
&AB-Flat: U(1,2)&Jeffrey & 499 & 9.06 & 9.08 & (481, 517) \\
& &Poi($\hat{N}_{M_b}$)& 495 & 8.15 & 9.61 & (478, 511) \\
&  &Poi($\hat{N}_{Nour}$)& 499 & 8.86 & 8.98 & (480, 516) \\
P4(420)&Nour & - & 499 & 13.53 & 13.55 & (473, 523)\\
&Lee: U(1,2)&Jeffrey & 488 & 12.38 & 17.03 & (457, 531) \\
&AB-Flat: U(1,2)&Jeffrey & 511 & 17.20 & 20.50 & (481, 543) \\
& &Poi($\hat{N}_{M_b}$)& 489 & 12.81 & 16.84 & (463, 511) \\
&  &Poi($\hat{N}_{Nour}$)& 505 & 15.47 & 16.32 & (478, 533) \\
& & &  &  &  & \\
\multicolumn{7}{|c|}{when no directional knowledge on $\phi$ is available}\\
P1&AB-Flat: U($\hat{c}$,2)&Jeffrey & 433 & 12.57 & 67.88 & (410, 459) \\
& &Poi($\hat{N}_{M_b}$)& 447 & 16.33 & 55.68 & (416, 481) \\
P2&AB-Flat: U($\hat{c}$,2)&Jeffrey & 448 & 10.24 & 53.38 & (428, 466) \\
 & &Poi($\hat{N}_{M_b}$)& 450 & 10.68 & 51.28 & (428, 470) \\
P3&AB-Flat: U($\hat{c}$,2)&Jeffrey & 474 & 6.63 & 26.67 & (460, 486) \\
& &Poi($\hat{N}_{M_b}$)& 473 & 6.52 & 27.64 & (459, 485) \\
P4&AB-Flat: U($\hat{c}$,2)&Jeffrey & 458 & 10.56 & 43.66 & (436, 477) \\
& &Poi($\hat{N}_{M_b}$)& 452 & 9.75 & 49.07 & (432, 470) \\
\hline
\end{tabular}
\end{minipage}
\label{tab:recapture_prone}
\end{center}
\end{table}

Now we turn to the recapture averse cases and associated results are presented in Table $\ref{tab:recapture_aversion}$. Population P5 and P6 in Table $\ref{tab:recapture_aversion}$ demonstrates a case of recapture averse population with $p_{1\cdot}<p_{\cdot1}$. For P5, Bayes estimate corresponding to Jeffrey's prior performs moderately for $U(\hat{c}, 1)$ and if it is not known that $\phi$ is less than 1, then the other priors $U(\hat{c}, 2)$ overestimate the \emph{N} due to low capture probabilities. Poisson prior with $\lambda=\hat{N}_{M_b}$ also misdirect the estimator due to same reason. In case of moderately high capture probabilities with $p_{1\cdot}<p_{\cdot1}$ in P6, our proposed strategy with $U(\hat{c}, 1)$ performs very well and other two estimates are also reasonably good. With the availability of the knowledge that $\phi$ is less than 1, Jeffrey's prior is relatively a better choice than poisson. Population P7 considers high capture probabilities with $p_{1\cdot}>p_{\cdot1}$. Estimate corresponding to the prior limit ($\hat{c}, 1$) is better than other two priors. Though these other two estimates can be considered as good if we ignore their slight overestimation. For population P8 also, prior $U(\hat{c}, 1)$ with Jeffrey's prior on \emph{N} produces reasonably good estimate whereas the other two priors are highly overestimates as the second capture probability is very small. For the situation when $p_{1\cdot}>p_{\cdot1}$, we recommend the use of poisson prior when no directional information on $\phi$ is available. Overall results from Table $\ref{tab:recapture_aversion}$ indicate that our approximate Bayes estimate with prior $U(\hat{c}, 1)$ for $\phi$ works very well but one can use this range only when it is known that $\phi<1$. If such information on $\phi$ is not available, then other two priors can be employed with poisson prior for \textit{N}. The results in these four tables also tell us that proposed approach, from other two prior limits, works reasonably good for high capture probabilities. It is also noted that estimates from Pois($\lambda=\hat{N}_{M_b}$) prior have smaller RMSE than that of Jeffrey's for high List-1 capture probability. Prior $\pi(\phi)=U(\hat{c}, 2)$ performs satisfactory for large $p_{1\cdot}$, without considering the fact that $\phi<1$ and in that situation, $\pi(N)=$ Poi($\hat{N}_{M_b}$) works better than Jeffrey's.
\begin{table}[htbp]
\footnotesize
\begin{center}
\caption{Summary results reflecting the performances of the proposed approach and other relevant method applied to the populations P5-P8. Upper panel refers the situation when the knowledge, $\phi<1$, is available. Lower panel refers when nothing is available on $\phi$.}
\begin{minipage}{13cm}
\begin{tabular}{|lllcccc|}
\hline
Population& & Prior &  Average & Sample & Sample &  \\
(E($x_0$)) & Method: $\pi(\phi)$ & $\pi(N)$ & Estimate & SE & RMSE & $95\% CI$ \\
\hline
\hline
\multicolumn{7}{|c|}{when it is known that $\phi<1$}\\
P5(430)&Lee: U(0.2,1.4)\footnote{This prior is used as Lee et al.'s strategy fails to generate samples from U(0.2, 1), when $\phi>1$}& Jeffrey & 456 & 18.80 & 47.88 & (432, 505) \\
&AB-Flat: U($\hat{c}$,1)& Jeffrey & 482 & 12.39 & 21.62 & (461, 508) \\
& &Poi($\hat{N}_{M_b}$) & 542 & 69.93 & 81.45 & (488, 676) \\
P6(459)&Lee: U(0.2,1.4)$^a$& Jeffrey & 481 & 7.40 & 20.51 & (460, 528) \\
&AB-Flat: U($\hat{c}$,1)& Jeffrey & 495 & 8.38 & 9.98 & (480, 510) \\
& &Poi($\hat{N}_{M_b}$)& 504 & 11.38 & 12.03 & (485, 526) \\
P7(483)&Lee: U(0.2,1.4)$^a$& Jeffrey & 504 & 6.90 & 7.69 & (484, 535) \\
&AB-Flat: U($\hat{c}$,1)& Jeffrey & 500 & 5.00 & 5.01 & (490, 508) \\
& &Poi($\hat{N}_{M_b}$)& 499 & 4.97 & 5.07 & (489, 508) \\
P8(446)&Lee: U(0.2,1.4)$^a$& Jeffrey & 523 & 20.26 & 30.73 & (466, 589) \\
&AB-Flat: U($\hat{c}$,1)& Jeffrey & 483 & 8.72 & 19.31 & (468, 498) \\
& &Poi($\hat{N}_{M_b}$)& 481 & 8.90 & 21.08 & (466, 497) \\
& & & &  &  & \\
\multicolumn{7}{|c|}{when no directional knowledge on $\phi$ is available}\\
P5&AB-Flat: U($\hat{c}$,2)& Jeffrey & 534 & 21.25 & 40.33 & (497, 581) \\
& &Poi($\hat{N}_{M_b}$) & 609 & 108.33 & 153.77 & (523, 823) \\
P6&AB-Flat: U($\hat{c}$,2)& Jeffrey & 526 & 13.56 & 28.98 & (502, 550) \\
& &Poi($\hat{N}_{M_b}$)& 529 & 16.80 & 33.54 & (495, 539) \\
P7&AB-Flat: U($\hat{c}$,2)& Jeffrey & 513 & 6.50 & 14.61 & (501, 525) \\
& &Poi($\hat{N}_{M_b}$)& 507 & 5.92 & 9.29 & (497, 518) \\
P8&AB-Flat: U($\hat{c}$,2)& Jeffrey & 522 & 20.17 & 29.77 & (494, 562) \\
& &Poi($\hat{N}_{M_b}$)& 504 & 16.51 & 17.07 & (481, 538) \\
\hline
\end{tabular}
\end{minipage}
\label{tab:recapture_aversion}
\end{center}
\end{table}

This analysis suggests that when directional information on $\phi$ is unavailable, it is not possible to have reasonably good estimate for $N$ from flat objective prior on $\phi$ except in case of $x_{1\cdot}>x_{\cdot1}$ under $\phi<1$.  Moreover, subjectiveness makes an estimate prior sensitive which is not at all desirable. Hence, this is an challenging trade-off job and we formulate an modified approach for unavailable directional knowledge of $\phi$ in the next section.

\section{Solution when Directional Knowledge on $\phi$ is Unavailable}\label{sub:second methodology_combo}
\subsection{Methodology (AB-Con)}
\label{sub:second methodology}

In earlier section it is observed that when directional information on $\phi$ is not available, the reasonable choice [$\alpha,\beta$]=[$\hat{c},2$] generally outperforms other methods for human population, if one uses constant or unform prior on $\phi$. In practice, for human population, demographic or any other beneficiary type survey reflects a recapture prone nature among the individuals. But, if no such information is available, can we modify our earlier proposed strategy (in section \ref{sub:first methodology}) considering suitably minimum subjectiveness in prior selection on $\phi$ so that reasonably good estimate can be obtained? In this section we try to set an conjugate prior on $\phi$ and therefore, investigate how our Bayes estimates, based on different potential loss functions, performs if hyperparameters are empirically estimated.

Unlike the previous case, we suggest that prior on $\phi$ is dependent on \emph{N} and therefore, joint prior distribution becomes $\pi(\Theta)=\pi(p_{1\cdot})\pi(\phi|N)\pi(N)$. We support the argument made by Lee et al. (2003\cite{Lee03}) that, in practice, it is necessary to restrict the range of $\phi$ to be between some $\alpha$ and $\beta$. Let us restrict $\phi \in [\alpha,\beta]$ and consider a conjugate prior on $\phi$ as $\pi(\phi)=\mbox{GB-I}(a,b,1,rate=1/\beta)$, for given \emph{a}, \emph{b} and $\beta$. Hence, $\pi(\phi|\alpha,\beta,a,b) \propto \phi^{a-1}(1-\phi/\beta)^{b-1} \times \mathcal{I}_{[\alpha,\beta]}(\phi)$. Since, $c=\phi p$ and $c<1$, then $p^{-1}$ might be thought as a good choice for the upper limit of $\phi$, hence $\beta=p^{-1}$. In practice, $\hat{c}$ is taken as a good choice for $\alpha$ and $p$ can be obtained using the relation $p=x_{01}/(N-x_{1\cdot})$, suggested by Llyod(1994). This suggestion leads to a conditional posterior of $\phi$ as a well-known probability density function from which one can directly generate Gibbs samples. The conditional posterior density will be
\begin{eqnarray}
\pi(\phi|N,a,b)&\propto & \mbox{GB-I}(x_{11}+a,x_{10}+b,1,rate=1/\beta)\times \mathcal{I}_{[\hat{c},\beta]}(\phi),\label{cond_posterior_phi_4}
\end{eqnarray}
where $\beta=p^{-1}=(N-x_{1\cdot})/x_{01}$ and \emph{a} and \emph{b} are chosen by equating $E_{\pi}(\phi|\alpha,\beta,a,b)$ with $c/p=c\beta$ to maintain the inter-relationships among model parameters in prior selection. Since, $E_{\pi}(\phi|\alpha,\beta,a,b)=\beta a(a+b)^{-1}$ which implies $a(a+b)^{-1}=c$. We choose $a=t(x_{11}/x_0)$ and $q=t(x_{10}/x_0)$ where \emph{t} ($>0$) is a tuning parameter that regulates the variance of the prior density such that $V_{\pi}(\phi)=\textit{O}(t^{-1})$. Remaining parameters in $\Theta$, i.e. $p_{1\cdot}$ and $N$, have same prior setup as mentioned in section \ref{sub:first methodology}. Hence, we can perform a simple Gibbs sampling MCMC technique with conditional posterior of $\phi$ in (\ref{cond_posterior_phi_4}) and other two conditional posterior densities $\pi(N-x_0|\phi,p_{1\cdot})$ and $\pi(p_{1\cdot}|N)$ exactly same as in section \ref{sub:first methodology}. Here also the hyper-parameter $\lambda$ is replaced by $\hat{N}_{M_{b}}$ as before for poisson prior. Firstly, we fix initial values $p_{1\cdot}^{(0)}$ and $p^{(0)}$ and the prior variance tuning parameter $t$. $\phi^{(0)}$ is simulated from conjugate prior GB-I with initial $\beta^{(0)}=1/p^{(0)}$. Then generate $N^{(0)}$ from its posterior $\pi(N-x_0|p_{1\cdot},p)$ replacing $p_{1\cdot}$ and $p$ by $p_{1\cdot}^{(0)}$ and $p^{(0)}$ respectively. Then, subsequent steps in Gibbs sampling is carried out as follows.
\\
\\
\texttt{Step 1: Simulate $p_{1\cdot}^{(1)}$ and $\phi^{(1)}$ from $\pi(p_{1\cdot}|N^{(0)})$ and $\pi(\phi|N^{(0)},a,b)$, in (\ref{cond_posterior_phi_4})\\ respectively, where $\beta=1/p^{(0)}$. \\
Step 2: Obtain $p^{(1)}=\hat{c}/\phi^{(1)}$.\\
Step 3: Generate \emph{N} from $\pi(N-x_0|p_{1\cdot}^{(1)},\phi^{(1)})$.\\
Step 4: Repeat the above three steps until the convergence is reached.} \\
\\
Hence, the values $\{N^{(h)}: k<h\leq 2k\}$, where \emph{k} is the chosen burn-in period, are believed to be a very large sample from the resultant approximate posterior distribution $\pi(N|\underline{\textbf{x}})$. \emph{k} is chosen based on the performance of $\hat{R}^{1/2}$ as stated earlier. To obtain estimate of true population size (\emph{N}) from the resultant posterior, we consider some potential loss functions, such as squared error, absolute error and maximum a posteriori (MAP) loss functions which produce estimators respectively as posterior mean ($\hat{N}_{MEAN}$), median ($\hat{N}_{MED}$) and mode ($\hat{N}_{MAP}$). Casella (1986) suggested another estimate obtained by minimizing the squared relative error loss function $L(N,\hat{N})=\left( \frac{N-\hat{N}}{N} \right)^2$, and the corresponding estimate is $\hat{N}_{SRE}=E_{\pi(N|D)}(N^{-1})/E_{\pi(N|D)}(N^{-2})$. In practice, $\hat{N}_{SRE}$ is obtained using the ratio of $\sum{1/N^{(h)}}$ and $\sum{1/[N^{(h)}]^2}$ from posterior sample $\{N^{(h)}: k<h\leq 2k\}$. One feature of this setup is that though it uses informative prior but the hyper-parameters (\emph{a}, \emph{b}, $\lambda$) are taken as functions of data for given the variance tuning parameter \emph{t}. Hence, one can say this approximate Bayes computation is closely related to empirically Bayesian statistics. In the next section, we numerically evaluate the performance of aforesaid MCMC algorithm in order to estimate $N$ when directional knowledge on $\phi$ is absent, equivalently, domain knowledge is $\phi>0$. We call this method as \textit{AB-Con} method since conjugated subjective prior choice is made for $\phi$ when directional knowledge on $\phi$ is \textit{unavailable}.
\subsection{Numerical Illustrations}
\label{sub:second illustration}

\begin{figure}[ht]
\centering
\includegraphics[height=8in,width=5.5in]{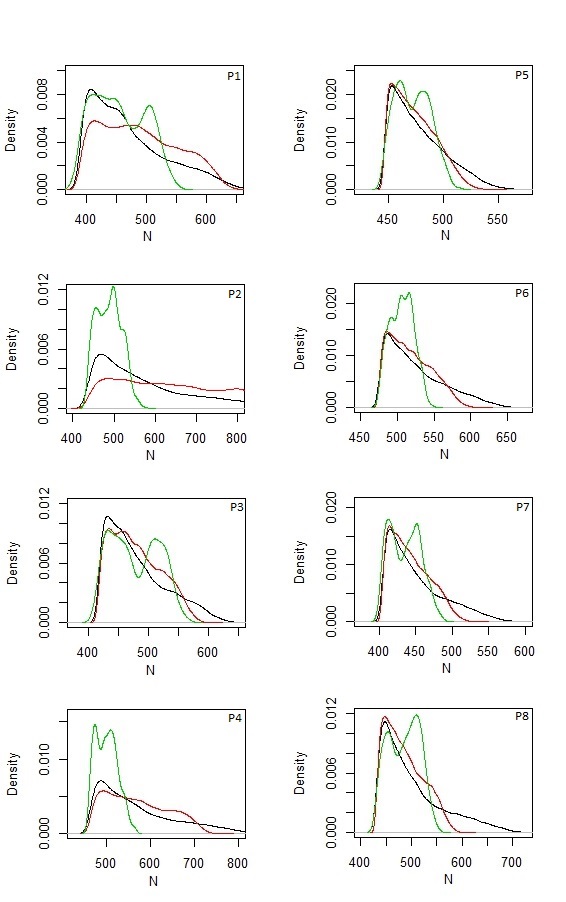}
\caption{Posterior distributions of \emph{N} based on AB-Con method (\textit{Black} and \textit{Red} lines respectively for Jeffrey's and Pois($\hat{N}_{M_b}$) priors for $N$) and Lee's method (\textit{Green} line).}
\label{fig:Fig01}
\end{figure}

Let us consider all the simulated populations discussed in section \ref{sub:first illustration} in order to illustrate the AB-Con method and also to suggest efficient priors under different loss functions considered. Prior belief on $\phi$ is considered with a reasonable value of $t=20$. Hence, we observe the performance of $\hat{R}^{1/2}$ for all populations and fix a general \emph{k} at 7000. Resulting posteriors from AB-Con method using Jeffrey's and Poisson priors on $N$ are shown in Figure \ref{fig:Fig01} along with posteriors from Lee et al. (2003\cite{Lee03}). Burn-in for Lee's(2003\cite{Lee03}) method is set at 150. Figure \ref{fig:Fig01} shows that Lee's method produces tighter posteriors than AB-Con but it is bimodal in nature for almost all cases. Posterior from Jeffrey's prior in AB-Con method is almost similar with poisson prior across populations but it has larger variability than poisson. In some cases, Lee(2003\cite{Lee03}) is better in terms of squared error or maximum-a-posteriori loss, but for bimodal
posteriors the higher mode is not close to the true value (as here, true $N$ is 500).

Final Estimates of \emph{N} are obtained by averaging over 200 posterior replications for each loss function. S.E. of each estimate is computed over 200 replicated estimates. It is clear from Table \ref{tab:Lossfunction} that poisson and Jeffrey's prior performs better respectively in case of squared error and absolute error loss. Specifically when $x_{1\cdot}<x_{\cdot1}$, Jeffrey's prior performs better for both the loss functions. MAP-based estimator significantly underestimates $N$ for recapture prone populations. Under this loss function, poisson prior is generally better than Jeffrey's. In particular for $x_{1\cdot}>x_{\cdot1}$, both the priors have similar performance. $\hat{N}_{SRE}$ would be effective in general with Jeffrey's prior for $N$. Indeed, $\phi>1$ corresponds to the most likely case in human demographic studies. For example, a specialised survey is conducted following a large census count, e.g. Post Enumeration Survey (PES), $x_{1\cdot}<x_{\cdot1}$ is often experienced. For sensitive dual survey, e.g. estimation of drug abused population size, $x_{1\cdot}>x_{\cdot1}$ is usually observed for time ordered samples. Hence, in order to find an estimator when no information on $\phi$ is available, Pois($\lambda=\hat{N}_{M_b}$) is suggested as a reasonably good selection for $\pi(N)$ if maximum-a-posteriori loss function is the objective of choice. Otherwise, for posterior median and Casella (1986) suggested loss function squared relative error (SRE), Jeffrey's non-informative prior on $N$ is preferred.
\begin{table}[ht]
\footnotesize
\begin{center}
\caption{Estimates of \emph{N} by \textit{AB-Con} method from different loss functions, Lee's(2003) and their associated s.e. in ( ) for all simulated populations, when directional knowledge on $\phi$ is NOT available.}
\begin{minipage}{15cm}
\begin{tabular}{|llccccc|}
\hline
  & Prior & & & & & \\
Popln. & $\pi(N)$ &Lee\footnote{Prior $\pi(\phi)=U(0.5,2)$ is chosen by the \textit{trial-and-error} method discussed in Lee et al. (2003[18])} & $\hat{N}_{MEAN}$& $\hat{N}_{MAP}$ & $\hat{N}_{MED}$ & $\hat{N}_{SRE}$\\
\hline
 &&&&&&\\
P1 & Jeffrey & 468 (20.56)&  483 (13.92)& 410 (10.52) & 467 (12.90)& 467 (12.51)\\
 & P($\hat{N}_{M_b}$) & - & 499 (23.34)& 419 (13.52) & 494 (22.77) & 483 (18.67)\\
P2 & Jeffrey & 483 (18.45)& 493 (12.04) & 433 (9.66) & 481 (11.13) & 483 (10.98)\\
 & P($\hat{N}_{M_b}$) & - & 489 (13.12)& 438(13.22)& 486 (12.87) & 482 (12.02)\\
P3 & Jeffrey & 485 (6.61)& 495 (7.40) & 464 (6.45) & 488 (6.90) & 492 (7.06)\\
 & P($\hat{N}_{M_b}$) & - & 490 (6.94) & 463 (6.17) & 486 (6.81) & 488 (6.69)\\
P4 & Jeffrey & 471 (8.11)& 473 (10.81) & 429 (8.61) & 461 (9.85) & 466 (9.94)\\
 & P($\hat{N}_{M_b}$) & - & 463 (9.91) & 429 (8.62) & 458 (9.69) & 458 (9.57)\\
 & &&&&&\\
P5 & Jeffrey & 474 (20.80)& 566 (14.85) & 454 (10.33) & 533 (12.13) & 530 (11.30)\\
 & P($\hat{N}_{M_b}$) & - & 660 (18.50) & 498(10.12) &  646(17.36) &  595(9.45)\\
P6 & Jeffrey & 512 (15.76) & 558 (11.57) &478 (7.63) & 544 (9.55) & 544 (8.97)\\
 & P($\hat{N}_{M_b}$) & - & 570 (19.89)& 490 (14.12) & 561 (18.96) & 553 (15.09)\\
P7 & Jeffrey & 516 (6.17)& 539 (7.62)& 492 (5.45) & 526 (6.67) & 532 (6.72)\\
 & P($\hat{N}_{M_b}$) & - & 528 (6.86)& 493 (5.16) & 524 (6.55) & 525 (6.36)\\
P8 & Jeffrey & 517 (13.02)& 525 (10.58)& 460 (7.36) & 505 (9.16) & 510 (8.98)\\
 & P($\hat{N}_{M_b}$) & - & 507 (9.45)& 462 (8.13) & 501 (9.35) & 501 (8.83)\\
& &&&&&\\
\hline
\end{tabular}
\end{minipage}
\label{tab:Lossfunction}
\end{center}
\end{table}

\section{Real Data Application}
\label{sec:real data}

Greenfield (1975) reports some of the results of a Population Change Survey in Malawi, conducted by the National Statistical Office between 1970 and 1972, to estimate birth, death and migration rates. The sample was stratified into five areas: Blantyre, Lilongwe and Zomba urban areas; other urban areas and rural areas. However, in this article we confine our attention to data on death records only for two strata - \textit{Lilongwe} ($\hat{c}=0.593$) and \textit{Other urban areas} ($\hat{c}=0.839$) based on different kind of $\hat{c}$. Nour (1982 \cite{Nour82}) estimated these death sizes as \textit{378} and \textit{3046} for \textit{Lilongwe} and \textit{Other urban areas} respectively assuming the fact that two data sources are positively correlated (i.e. $\phi>1$) in a human demographic study. To implement our both methods, 200 parallel chains are generated from different randomly selected starting points for all the MCMC methods discussed earlier sections (i.e. for AB-Flat, AB-Con and Lee's(2003) method). Then, we compute $\hat{R}^{1/2}$ (with respect to \emph{N}) to determine the burn-in period \emph{k}. In Table \ref{tab:realresult}, upper and lower panels respectively represent the results corresponding to \textit{Lilongwe} and \textit{Other Urban Area}.

At first, we analyse the data assuming that the two populations are recapture prone as in human demographic study, positive list-dependence is often occurred. So, we also use the poisson prior with $\lambda=\hat{N}_{Nour}$ in addition to other two priors - Jeffrey's and Pois($\lambda=\hat{N}_{M_{b}}$), for method AB-Flat. In the first half of both panels of the Table \ref{tab:realresult}, for the prior \emph{U}(1, 2) on $\phi$, first, second and third row correspond to the Jeffrey's, Pois($\lambda=\hat{N}_{M_b}$) and Pois($\lambda=\hat{N}_{Nour}$) priors respectively. We fix burn-in \textit{k} generally at 500 for Lilongwe and 3000 for Other urban areas (see Figure 2) and record
the remaining \textit{k} values in each of 200 chains. Proposed AB-Flat approach with any suitable prior gives very close and efficient results in comparison to Nour's. Moreover, it is also found that in \emph{Lilongwe}, peoples are more keen to capture the death records again than that of survey time compared to \emph{Other urban areas}. Our estimate with the assumption $\phi>1$ indicates that around 380 deaths occurred in \emph{Lilongwe} and around 3030 deaths occurred in \emph{Other urban areas}. As \emph{N} is large, we also examine that effect of chosen larger $\beta$ for $\pi(\phi)$, on the estimate becomes less.
\begin{figure}[h]
\centering
\includegraphics[width=5.5in]{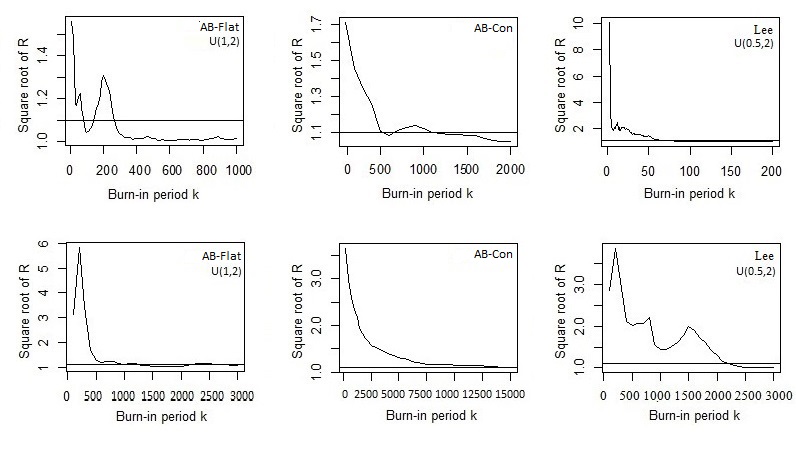}
\caption{Plot of $\hat{R}^{1/2}$ against burn-in period \emph{k} for MCMC with Jeffrey's prior for $N$ in each of AB-Flat, AB-Con and Lee's methods. First and second rows are for \emph{Lilongwe} and \emph{Other urban areas} respectively. Horizontal line presents the threshold 1.1 for $\hat{R}^{1/2}$.}
\label{fig:Fig02}
\end{figure}
\begin{table}[h]
\footnotesize
\centering
\caption{Bayesian estimates of total number of deaths using AB-Flat, AB-Con and Lee's methods. s.e. is computed based on sample posterior distribution and the $95\%$ posterior credible intervals for \textit{N} and \textit{$\phi$} is determined based on percentile method.}
\begin{minipage}{15cm}
\begin{tabular}{|lllccccc|}
\hline
& & &  &  & $95\%$ CI &  & $95\%$ CI\\
Method & $\pi(\phi)$ & $\pi(N)$ & $\hat{N}$ & s.e.($\hat{N}$) & of \emph{N} & $\hat{\phi}$ & of $\phi$\\
\hline
& &  & & &  &  & \\
\multicolumn{8}{|c|}{Lilongwe} \\
& &  &  &  &  & & \\
\multicolumn{8}{|l|}{\underline{Consider $\phi>1$}} \\
AB-Flat & U(1, 2)& Jeffrey & 380 & 3.91 & (373, 388) & 1.35 & (1.19, 1.54)\\
& & P($\hat{N}_{M_b}$) & 374 & 2.57 & (370, 380) & 1.35 & (1.20, 1.53)\\
& & P($\hat{N}_{Nour}$) & 379 & 3.12 & (373, 385) & 1.35 & (1.20, 1.54)\\
& & &  &  &  &  & \\
\multicolumn{8}{|l|}{\underline{Consider $\phi>0$}\footnote{$\phi>0$ refers natural domain of $\phi$, as no directional knowledge is considered}} \\
AB-Flat & U($\hat{c}$, 2)& Jeffrey & 362 & 5.25 & (355, 374) & 0.94 & (0.75, 1.22)\\
& &P($\hat{N}_{M_b}$) & 361 & 4.03 & (354, 369) & 0.94 & (0.75, 1.19) \\
Lee & U(0.5, 2)\footnote{This prior is chosen based on \textit{trial-and-error} method discussed in Lee et al. (2003\cite{Lee03})} & Jeffrey & 354 & 5.80 & (348, 370) & 0.74 & (0.57, 1.14)\\
AB-Con & MEAN& Jeffrey & 366 & 4.90 & (357, 375) & 1.03 & (0.82, 1.24)\\
& &P($\hat{N}_{M_b}$) & 363 & 3.81 & (357, 371) & 1.03 & (0.82, 1.24)\\
& MED& Jeffrey & 362 & 5.37 & (354, 372) & 0.95 & (0.75, 1.22)\\
& &P($\hat{N}_{M_b}$) & 361 & 4.61 & (355, 371) & 0.95 & (0.75, 1.25) \\
& MAP& Jeffrey & 352 & 5.25 & (348, 366) & 1.04 & (0.62, 1.87)\\
& &P($\hat{N}_{M_b}$) & 353 & 5.93 & (348, 372) & 1.04 & (0.62, 1.87) \\
& SRE& Jeffrey & 365 & 4.56 & (357, 373) & 0.86 & (0.75, 1.01)\\
& &P($\hat{N}_{M_b}$) & 363 & 3.65 & (357, 370) & 0.86 & (0.75, 1.04) \\
& & &  &  &  &  & \\
\multicolumn{8}{|c|}{Other urban areas} \\
& &  &  &  &  & & \\
\multicolumn{8}{|l|}{\underline{Consider $\phi>1$}} \\
AB-Flat & U(1, 2)& Jeffrey & 3030 & 51 & (2973, 3172) & 1.12 & (1.06, 1.26)\\
& & P($\hat{N}_{M_b}$) & 3056 & 46 & (3000, 3180) & 1.12 & (1.06, 1.22)\\
& & P($\hat{N}_{Nour}$) & 3027 & 40 & (2978, 3133) & 1.16 & (1.06, 1.28)\\
& & & &  &  &  & \\
\multicolumn{8}{|l|}{\underline{Consider $\phi>0$}$^a$} \\
AB-Flat & U($\hat{c}$, 2)& Jeffrey & 2860 & 41 & (2816, 2967) & 0.94 & (0.89, 1.05)\\
& & P($\hat{N}_{M_b}$) & 2873 & 43 & (2826, 2985) & 0.94 & (0.89, 1.05)\\
Lee& U(0.5, 2)$^b$ & Jeffrey & 3455 & 223 & (3096, 3870) & 1.55 & (1.19, 1.97)\\
AB-Con & MEAN & Jeffrey & 3152 & 119 & (2951, 3381) & 1.24 & (1.03, 1.49)\\
& &P($\hat{N}_{M_b}$) & 3146 & 101 & (2972, 3348) & 1.24 & (1.04, 1.49)\\
& MED& Jeffrey & 3109 & 149 & (2907, 3441) & 1.20 & (0.99, 1.54)\\
& &P($\hat{N}_{M_b}$) & 3131 & 135 & (2932, 3416) & 1.20 & (0.99, 1.55) \\
& MAP& Jeffrey & 2953 & 215 & (2774, 3491) & 1.29 & (0.86, 1.95)\\
& &P($\hat{N}_{M_b}$) & 3056 & 263 & (2780, 3580) & 1.28 & (0.86, 1.94) \\
& SRE& Jeffrey & 3109 & 109 & (2935, 3344) & 1.13 & (1.00, 1.37)\\
& &P($\hat{N}_{M_b}$) & 3113 & 97 & (2956, 3320) & 1.14 & (0.99, 1.37) \\
& & &  &  &  &  & \\
\hline
\end{tabular}
\end{minipage}
\label{tab:realresult}
\end{table}

Bottom half of both the panels in Table \ref{tab:realresult} present the results for both the strata when no information is available on the nature of behavioral response effect, i.e. AB-Con method. We fix burn-in \emph{k} generally at 2000 for \emph{Lilongwe} and 15000 for \emph{Other urban areas} (\textit{see} Figure \ref{fig:Fig02}). Figure \ref{fig:Fig02} also suggests that burn-in period \textit{k} in Lee's method is 100 and 3000 respectively for the two strata. For poisson prior on $N$, the hyper-parameter $\lambda$ is replaced by $\hat{N}_{M_{b}}$. Prior \textit{U}($\hat{c}$, 2) in AB-Flat method says that the estimated number of deaths in \emph{Lilongwe} is 360 with a $95\%$ credible interval (357, 365) and in \emph{Other urban areas} is around 2856 with a $95\%$ credible interval (2828, 2929), which are both underestimation than respective Nour's estimates. These results disagree with the recapture proneness of the population assumed by Nour (1982). Without considering any directional knowledge on $\phi$ (i.e. AB-Con method) the posterior mean, median and SRE estimates suggest that number of deaths in \emph{Lilongwe} is around 363 and contradict the assumption made by Nour(\cite{Nour82}) that this population is \textit{recapture prone}. For \emph{Other urban areas}, our analysis agrees that this population is \textit{recapture prone} as $\phi$ is around 1.20 and corresponding estimate of $N$ is nearly 3130, which is greater than Nour's. For both strata, our MAP based Bayes estimates provide lower estimates. Lee(2003) highly overestimates the death size in \emph{Other urban areas} than all the proposed estimates including Nour's(1982).

\section{Summary and Conclusions}
\label{sub:conclusion}

In the present article, we have presented two approximate Bayesian approaches under a general framework for dual-record system (DRS) where behaviour response effect might play a significant role along with time variation effect. Here we suggest efficient approximate Bayesian computation strategies conditionally and unconditionally on the directional knowledge available on $\phi$. The first one is formulated with uniform prior whereas the second one depends on subjective conjugate prior based on structural relationships among the underlying parameters. Some features of the first approach with uniform prior on behaviour effect ($\phi$) are: noninformative prior for \emph{N} and $p_{1\cdot}$ is used and a reasonable range for $\phi$ is always available with or without the help of the available directional information on $\phi$. When $\phi <1$, specification of the lower bound of $\phi$ by $\hat{c}$ works successfully. But when $\phi >1$, our study concludes that estimate with $\pi(\phi)\propto\emph{U}(1, 2)$ and $\pi(N)\propto N^{-1}$ is expected to be superior than Nour's estimate in terms of smaller RMSE and reasonably better CI. Moreover, the upper limit $\beta$ is not at all influential if the nature of $\phi$ is correctly known. It is found that estimates from poisson prior with $\lambda=\hat{N}_{M_b}$ are less efficient than Jeffrey's. Hence, we conclude that the first approximate Bayes approach (discussed in section \ref{sub:first methodology}) performs very well based on the information on the possible range of underlying $\phi$, when available. In practice, experts can usually judge whether the specified population is either recapture prone (i.e. $\phi >1$) or averse (i.e. $\phi <1$) from past studies. If so, our strategy with noninformative prior on $\phi$ has significant improvement over Nour (1982 \cite{Nour82}) in terms of efficiency.

An alternative Bayes approach (discussed in section \ref{sub:second methodology}) with informative generalised beta prior is also proposed when there is no reliable information available on $\phi$. Some features of this approximate empirical type Bayes approach are the following: For $\phi <1$, MAP-based estimates are very efficient (compared to other loss functions) when the capture probabilities are high. In contrast, the other two estimates, $\hat{N}_{MED}$ and $\hat{N}_{SRE}$, with Jeffrey's prior or $\pi(N)\equiv$ Pois($\lambda=\hat{N}_{Nour}$) perform relatively better than $\hat{N}_{MAP}$ for $\phi >1$. When directional information is not available, $\hat{N}_{MAP}$ obtained from $\pi(N)\equiv$ P($\lambda=\hat{N}_{M_b}$) would be a unique choice among the two approaches. It is also found that the second approach improves the performance than that of first approach for known information of $\phi >1$. For recapture averse population ($\phi <1$), the first approach is little better because of its relatively tighter prior domain. Hence, our proposed methods can be used to have a better and easily computable estimate of population size from this complex dual system. Apart from the computational advantage, these methods are transparent and relatively easy to explain to the practitioner. Though our methods incorporate subjective elements through the choice of priors, as necessary, but this subjectiveness helps the underlying model to successfully get rid of the identifiability problem.

\section*{Acknowledgements}

This work is partially supported by the research fellowship award (No. $09/093(0125)/2010-EMR-I$) given to the first author from Council of Scientific and Industrial Research (CSIR), India. We are grateful to Dr. Sourabh Bhattacharya for his valuable comments and suggestions.













\end{document}